\documentclass{elsart}

\usepackage{natbib}

\usepackage{graphicx}
\usepackage{epsfig}

\usepackage{amssymb}

\begin{document}

\begin{frontmatter}



\title{Detecting gravitational waves from accreting neutron stars}


\author{Anna L. Watts\corauthref{cor}}
\address{Astronomical Institute ``Anton Pannekoek'', University of
  Amsterdam, Kruislaan 403, 1098 SJ Amsterdam, the Netherlands}
\corauth[cor]{Corresponding author}
\ead{A.L.Watts@uva.nl}
\ead[url]{http://www.astro.uva.nl/$\sim$awatts/}

\author{Badri Krishnan}
\address{Albert-Einstein-Institut, Max-Planck-Institut f\"ur Gravitationsphysik, Am
  M\"uhlenberg 1, 14476 Golm, Germany}

\begin{abstract}
The gravitational waves emitted by neutron stars carry unique
information about their structure and composition.  Direct detection
of these gravitational waves, however, is a formidable technical
challenge. In a recent study we quantified the hurdles facing
searches for gravitational waves from the known accreting neutron
stars, given the level of uncertainty that exists regarding spin and orbital
parameters.  In this paper we reflect on our conclusions, and issue 
an open challenge to the theoretical community to consider how
searches should be designed to yield the most astrophysically
interesting upper limits. With this in mind we
examine some more optimistic emission scenarios involving spin-down,
and show that there are technically feasible searches, particularly
for the accreting millisecond pulsars, that might place
meaningful constraints on torque mechanisms. We finish with a brief
discussion of prospects for indirect detection.  
\end{abstract}


\end{frontmatter}


\section{Introduction}
\label{intro}

The last few years have seen the commissioning of an advanced and
highly sensitive network of interferometric gravitational wave
detectors:
LIGO\footnote{http://www.ligo.caltech.edu}, 
GEO-600\footnote{http://www.geo600.aei.mpg.de},
VIRGO\footnote{http://wwwcascina.virgo.infn.it} and
TAMA\footnote{http://tamago.mtk.nao.ac.jp}.  These
observatories are now at or very close to their initial design
sensitivity, and order of magnitude improvements in performance are
expected over the next few years as they are upgraded to
incorporate major technological advances.  

There is no question that we are on the verge of the first direct
detection of these elusive ripples in space-time.  When this happens,
it will be a stunning confirmation of one of the main predictions of
the Theory of General Relativity.  It will also open up uncharted
territory for astrophysicists.  A gravitational wave can pass almost
unattenuated through matter that would absorb and re-emit a photon
many millions of times.  Gravitational waves will let us study
regions of the Universe that have until now remained hidden,
illuminating completely new physical processes.

The challenges involved, however, are immense.  The amplitudes from
even the strongest astrophysical sources are so weak that extricating
signals from the noise requires advanced search techniques and, where
feasible, signal templates.  When source parameters are poorly
constrained, searches rapidly become both computationally and
statistically untenable.  It is here that information from
electromagnetic astronomy can improve the efficiency of searches, and
make the difference between detecting or not detecting a source. 

A major target for current searches is the detection of the continuous
periodic signals expected from spinning neutron stars.  Emission
requires some ellipticity on the star, and theorists have been highly
inventive in conceiving of mechanisms that might generate a
quadrupole.  Possibilities include crustal mountains
\citep{Bildsten98, Ushomirsky00, Melatos05, Haskell06, Payne06, Vigelius08},
internal magnetic deformation \citep{Katz89, Cutler02, Haskell08}, and
internal r-mode oscillations \citep{Andersson99, Levin99, Andersson00, Andersson02,
  Heyl02, Wagoner02, Nayyar06, Bondarescu07}.  Angular momentum loss
via gravitational wave emission is also an attractive way of keeping
neutron star spin below the break-up limit of $\sim 1$ kHz
\citep{Lattimer07}.  Accretion should spin stars up to this limit
\citep{Cook94}, but both the accreting neutron stars and the
millisecond radio pulsars (their supposed progeny) spin far more
slowly \citep{Hessels06}.  The reason for this discrepancy might very
well be gravitational wave emission, as discussed below.

\section{The spin balance scenario}

Motivated by the idea that gravitational wave torque could balance
accretion torques, \citet{Bildsten98} pointed out that accreting
neutron stars might be steady beacons of gravitational waves.  In the
simplest scenario, where a neutron star (mass $M$, radius $R$, 
distance $d$)
has a quadrupole 
moment $Q$ that is stationary in the rotating frame of the star,
gravitational waves are emitted at a frequency $\nu = 2 \nu_s$, where
$\nu_s$ is the spin frequency. We assume an accretion torque 

\begin{equation}
N_a \approx \dot{M}\left(GMR\right)^{1/2}.
\end{equation}
(where accretion rate $\dot{M} = 4\pi R d^2 F/GM$ can be estimated from
the bolometric flux $F$), and a gravitational
wave torque 

\begin{equation}
N_\mathrm{gw} = - \frac{4\pi c^3 \d^2 \nu_s h_0^2}{5 G}
\end{equation}
where the gravitational wave amplitude $h_0$ associated with the
quadrupole $Q$ is defined as in
\citet{Jaranowski98}.  The assumption of spin balance then gives

\begin{equation}
h_0 = 3\times 10^{-27} \left(F_\mathrm{-8}\right)^{1/2} \frac{R_{10}}{M_{1.4}}
\left(\frac{ 1~\mathrm{kHz}}{\nu_s}\right)^{1/2}
\end{equation}
where $F_{-8} = F/10^{-8}$ ergs/cm$^2$/s, $R_{10} = R/10$ km and $M_{1.4} =
M/1.4M_\odot$.  The stronger the flux, the stronger the predicted
gravitational wave signal.

Whether this signal is detectable depends on a number of factors.
Firstly, the predicted signal must of course exceed the detectability
threshold.  This depends on the sensitivity of the detector, and the
detection level set by the detector teams (in simplistic terms,
one requires the amplitude to exceed that generated by noise processes
with a specified degree of probability).  However there
are other factors at play.  In all of the emission scenarios,
gravitational wave emission is related to the spin-frequency of the
neutron star.  This means that in order to fold long stretches of
data (necessary for such weak signals), we require an ephemeris for the
spin and orbital parameters.  If these parameters are known precisely,
only a single template must be searched.  If there is any uncertainty,
multiple templates must be searched, impacting both statistics (more
templates means more trials, so more chance of throwing up a spurious
result) and computational load (which sets feasible integration times).

In a recent study we assessed this scenario for all known accreting
neutron stars with some indication of spin rate, generating a
realistic assessment of detectability using accurate noise curves and
detection statistics for current and planned detectors \citep{Watts08}.
We included in our study the accreting millisecond pulsars (spins and
orbital parameters mostly well-constrained), the burst oscillation
sources (spins known to within a few Hz, orbits not so well
constrained), and the twin kHz QPO sources.  Whether the spin of the latter
class of sources is actually constrained at all by the separation of
the twin kHz QPO frequencies is
still a matter of debate.  However theoretical modeling indicates that there may be some link, and so
we included them in our study.   

\begin{figure} 
\begin{center}
\includegraphics[width=\textwidth, clip]{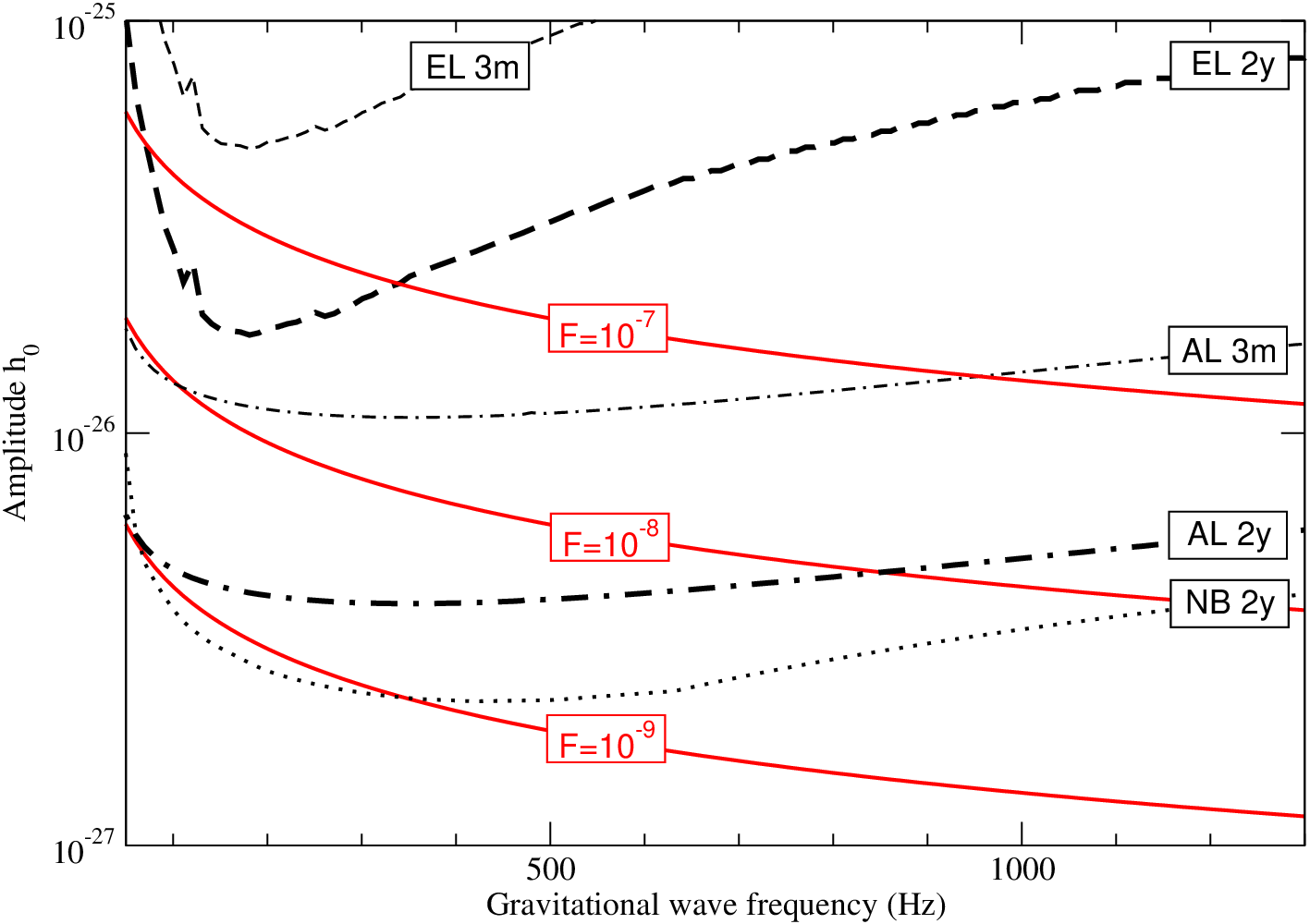}
\end{center}
\caption{Detectability for the spin balance scenario, with
  gravitational wave emission at twice the spin frequency of the
  neutron star, assuming that only one template need be searched (we
  have also averaged over all sky-positions and pulsar orientations,
  and chosen thresholds corresponding to a false alarm rate of 1\% and
  a false dismissal rate of 10\%).  The solid (red) lines show the
  predicted signal strength for a given X-ray flux $F$ in units of
  ergs/cm$^2$/s.  The broken (black) lines show detectability curves
  for various detector configurations (EL - Enhanced LIGO, AL -
  Advanced LIGO broad band, NB - Advanced LIGO narrow band envelope)
  and integration times (3m - 3 months, 2y - 2 years).  Enhanced LIGO
  is the configuration that will operate during the next LIGO science
  run in 2009-2010.  Advanced LIGO
  (http://www.ligo.caltech.edu/advLIGO) is the configuration that will
  operate a few years later after a major upgrade.  In each case we
  have assumed a coherent combination of data from two instruments.
  For the advanced detectors, we might very well have four instruments
  (three LIGO and one Virgo) operating in coincidence.  This would
  improve sensitivity by an additional factor of $\sqrt{2}$.}
\label{f1}
\end{figure}

\begin{figure} 
\begin{center}
\includegraphics[width=\textwidth, clip]{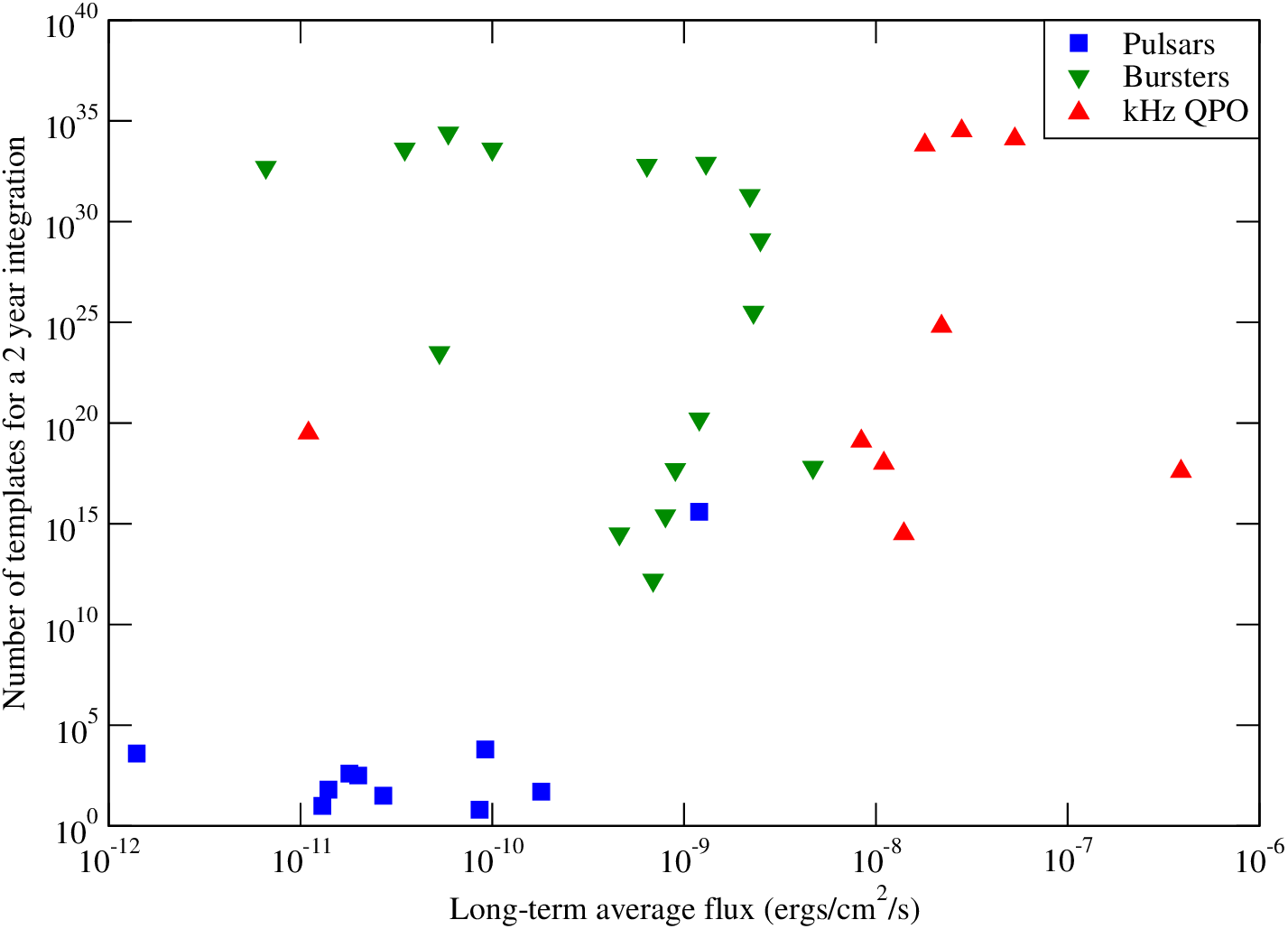}
\end{center}
\caption{Factors affecting detectability for the known accreting
  neutron stars where there is some indication of spin rate: the
  accreting millisecond pulsars, the burst oscillation sources, and
  the twin kHz QPO sources.  The plot shows the long-term
  time-averaged fluxes (compare to the detectability curves in Figure
  \ref{f1}), and the number of templates that would have to be
  searched for a 2 year integration, given the current uncertainty in
  spin and orbital parameters for each source (Figure \ref{f1} assumes
  that only a single template is searched) The best constrained
  sources tend to be those with the weakest predicted signals (in the
  spin balance scenario).  For reference, we note that
  ${\cal O}(10^{15})$ templates is \emph{very roughly} the largest
  number of templates that could be searched today using about a
  year's worth of LIGO data, and large (but still feasible) amounts of
  computational resources. }
\label{f2}
\end{figure}

Figure \ref{f1} shows the best case detectability for sources with a
given flux, assuming that all
spin and orbital parameters are sufficiently well constrained that
only one template need be used.  Figure \ref{f2} shows typical
long-term time-averaged fluxes
for the various sources (values from Table 1 of \citealt{Watts08}).
Comparing to Figure \ref{f1} we can see that the accreting millisecond
pulsars are simply too weak to be detectable, as are most of the burst
oscillation sources (given their frequencies, which tend to be
high)\footnote{The response times of the gravitational wave torque mechanisms
  are not well understood.  For systems that are transient accretors,
  like the accreting millisecond pulsars and many of the burst
  oscillation sources,
  it may be better to consider the flux in outburst rather than the
  long-term time-average flux.  This scenario, discussed in detail in
  \citet{Watts08} is somewhat more optimistic for some sources, since the
  increase in flux compensates for the short (typically
  $\sim$ months) outburst times.  It is not enough however to render
  any additional sources detectable by Advanced LIGO.}. The only sources that are bright enough to be detectable are
the kHz QPO sources.  Unfortunately, as is also clear from Figure
\ref{f2}, these are also the sources with the greatest number of
templates to be searched (for full details of the template estimates
see \citealt{Watts08}).   It turns out that the effect on statistics
alone, due to the increased number of trials, is enough to 
push most of the sources below the detectability threshold for
Advanced LIGO. Computational considerations impact things still
further by reducing 
feasible integration times.  If we search the full frequency band
implied by the separation of the twin kHz QPO frequencies, then only
one source remains above the detectability threshold for Advanced
LIGO.

The major factor hampering searches for
gravitational waves is clearly poor knowledge of spin and orbital parameters
for the brighter sources.  Efforts to detect accretion-powered
pulsations from these 
sources have so far been unsuccessful, and theoretical modeling
\citep{Kulkarni08} suggests that channeled accretion may simply not
occur at the highest accretion rates.  Several of the sources exhibit
Type I X-ray bursts, and it is possible that burst oscillations
(nuclear-powered pulsations seen only during bursts) may yet be
detected.  It is also possible that theoretical studies may eventually
reveal a firm link between kHz QPO frequency and spin.  However it is
perhaps time to consider more radical ways of obtaining spin
constraints.   

One possibility, for example, would be to follow up on
the modeling of \citet{Spitkovsky02}, which suggested a link between
the rise times of Type I X-ray bursts and the spin rate of the
star.   The rise time of the lightcurve from a Type I X-ray
  burst depends on the latitude at which ignition occurs, the
  speed $v_f$ at which the flame front propagates across the star, and
  the star's inclination with respect to the observer.
  \citet{Spitkovsky02} modeled flame spread across the
  surface of a rapidly rotating neutron star and showed that flame velocity $v_f \approx (gh)^{1/2}/(ft_n)^k$ where $g$ is the local gravitational
  acceleration, $h$ the scale height of the burning ocean, $t_n$ the
  timescale for nuclear burning and $f = 4\pi\nu_s \cos\theta$ the
  Coriolis parameter ($\theta$ is the latitude).  The power $k$
  depends on the degree of 
  frictional coupling between the top and bottom of the burning ocean,
  and is constrained to lie between 1/2 and 1.   Burst rise time
  therefore depends, at least in part, on spin rate.  The other
  parameters can in many cases be constrained, by theory or observation (burst shape, for example, can
 pinpoint ignition latitude, \citealt{Maurer08}).  Isolating the
 effects of spin from the 
 variation due to the other factors will still not be easy, but may
 be possible
 given the extensive archives of burst data that now exist. 
If the \citet{Spitkovsky02} picture survives more rigorous modeling
and test, then the 
fastest burst rise time should place at least an upper limit on the
spin rate of the star.  

Ideas like this, which are clearly more
speculative, are certainly worth exploring given the potential payoff
for the gravitational wave searches.   If we could measure spin to
within $\pm$ 5 Hz then four sources 
would move above the detectability threshold of Advanced LIGO 
provided that computational restrictions can be overcome.  This is
not an unreasonable hope:  several ways of improving the computational
situation were outlined in \citet{Watts08}.  For the other sources
better knowledge of spin alone is not sufficient and our knowledge of
the orbital parameters must also improve.  Fortunately there has
been major progress in this area in recent years using optical/IR observations
(for an overview see \citealt{Cornelisse08}).

\section{Moving beyond spin balance}

The legitimacy of the spin balance scenario, particularly on short
timescales, has always been a matter of debate.  Accretion luminosity
is observed to vary, and accretion episodes in many sources are
transient.  The timescales on which the various quadrupoles respond
are also very poorly understood - within the r-mode scenario, for
example, spin-down may be episodic.  

Scenarios where the source is out of balance, and where gravitational
wave torques might (at least temporarily) exceed accretion torques are
attractive, as predicted signal could then be larger.  The downside is
that one must then take additional search parameters into account in
the searches.  We will start however by considering the best case
scenario, assuming that we can track both spin and orbital parameters
so that we only have to search a single template.  One can write the
change in spin rate $\dot{\nu_s}$ in terms 
of the moment of inertia $I$ and the other parameters defined previously.

\begin{equation}
I_{45}\dot{\nu}_s =
1.4\times 10^{-14} ~\mathrm{Hz/s}~ d_\mathrm{kpc}^2\left[F_{-8}\left(\frac{
    R_{10}^3}{M_{1.4}}\right)^\frac{1}{2} - 11 \left(\frac{\nu_s}{1
      ~\mathrm{kHz}}\right) \left(\frac{h_0}{
      10^{-26}}\right)^2\right]
\label{spindown}
\end{equation}
where $I_{45} = I/10^{45}$ g cm$^2$ and $d_\mathrm{kpc} = d/1$ kpc.   Below a
certain flux the accretion term can be neglected and gravitational
wave induced spin-down is the dominant term.  Note that this equation
neglects any
other negative torques that may be present, such as those generated by
magnetosphere/disk interaction, jets, or magnetic dipole
spin-down. This means that we are considering the most 
optimistic scenario for gravitational wave emission.  

\begin{figure} 
\begin{center}
\includegraphics[width=\textwidth, clip]{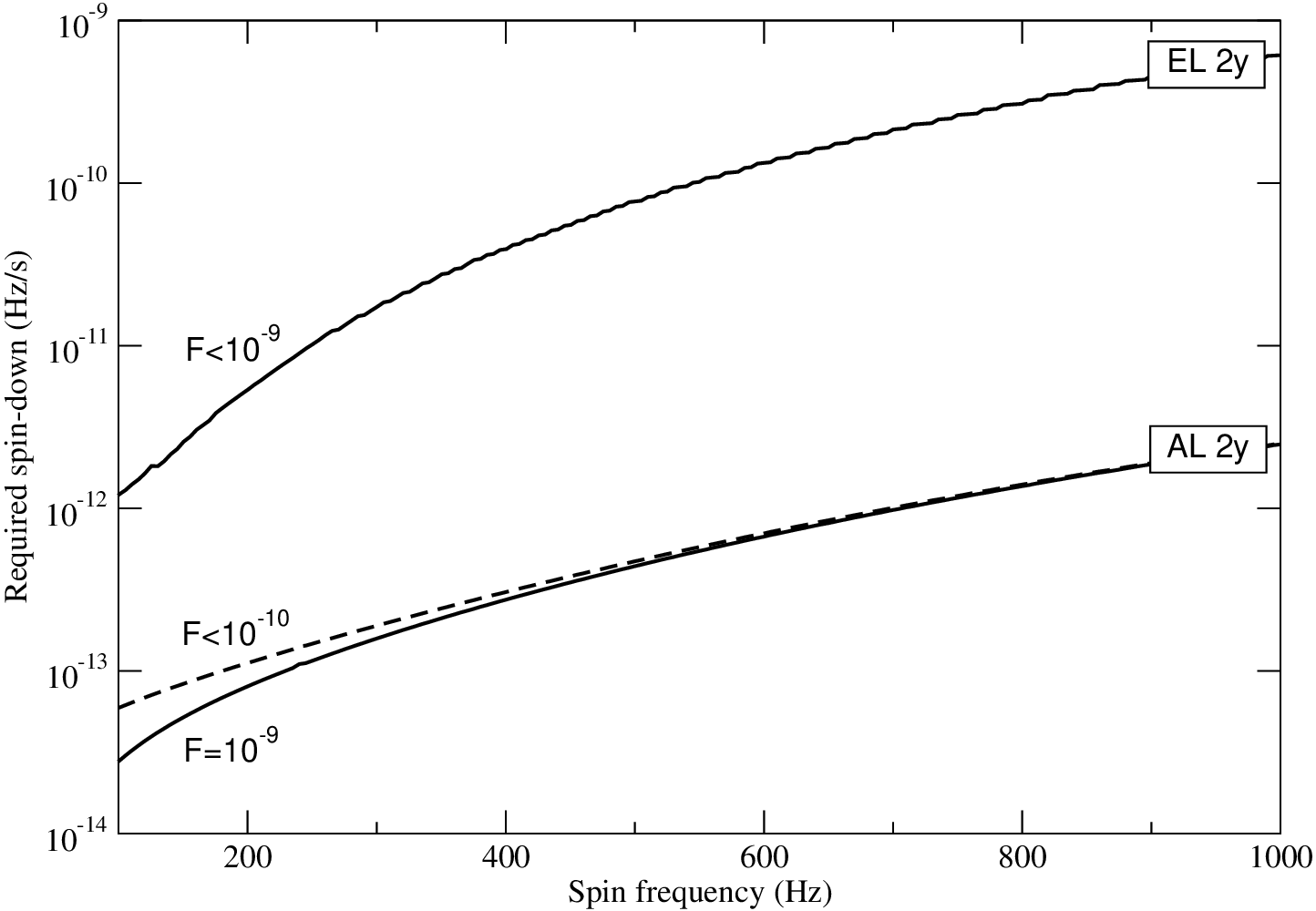}
\end{center}
\caption{For a signal to be detectable in the spin varying scenario,
  gravitational wave torque must exceed the accretion torque.  This
  means that for a signal to be detectable there must be an associated
spin-down (Equation \ref{spindown}).  This plot shows the required spin-down
for sources with low fluxes $F$ (in units of ergs/cm$^2$/s) to be
detectable in a 2 year integration 
by either Enhanced LIGO (EL) or Advanced LIGO (AL), assuming a
typical source
distance of 5 kpc.  As in Figure \ref{f1}, this is the best case
scenario, as it assumes perfect knowledge of spin and orbital parameters. }
\label{f3}
\end{figure}

To investigate the spin-down that would be associated 
with a detectable signal, we set $h_0$ equal to one of the
detectability curves plotted in Figure \ref{f1} (assuming a particular
detector configuration and an integration time).  Figure \ref{f3}
shows the spin-down rate that would be associated with such a signal,
for a typical source distance of $d_\mathrm{kpc} = 5$.  For a 2 year
integration, one would require $|\dot{\nu}_s| \sim 10^{-15}$ Hz/s
for there to be no additional template requirement.  For both
Enhanced and Advanced LIGO, however, a detectable signal from a low flux
source would have a much larger spin-down.  This means that it will be
necessary to search spin derivative  
templates if spin cannot be tracked precisely (for burst
oscillation sources, for example, or sources that are in quiescence).
For the sources where we cannot track spin precisely we would also
need to consider the likelihood of the source being in a spin-down
episode when observed.  Given this additional effect on detectability
we will concentrate for now on the accreting millisecond pulsars.  

We can compare the `detectable' values of spin-down to the 
spin derivatives that have been reported in the literature for the
accretion-powered millisecond 
pulsars (summarized in \citealt{Watts08}).  Pulse time of arrival
analysis has led to claims of both spin-up and spin-down, with upper limits on $|\dot{\nu}_s| \sim
10^{-12}$ Hz/s.  Timing analysis of these systems is a complex
business due to the extreme pulse profile variability, and there is
ongoing debate as to whether these spin derivatives are genuine or an
artifact \citep{Hartman08}.  However, a concerted effort is
underway to resolve these issues, and there is some hope of getting
reliable values of spin derivatives in the near future. 

If a spin-down larger than the detectability limit can be confirmed
for any of the accreting millisecond pulsars,
then simultaneous X-ray timing and gravitational wave searches would confirm
whether gravitational waves provide the limiting torque.   Given
current schedules, there will be a gap in simultaneous X-ray timing as
RXTE is scheduled to go offline in September 2009, in the middle of
the Enhanced LIGO 
science run, and ASTROSAT\footnote{http://meghnad.iucaa.ernet.in/$\sim$astrosat}, an Indian X-ray timing
mission, may not launch until 2010. However once Advanced LIGO starts
to operate, ASTROSAT should be
available to provide timing support.  Certainly by the time
that third generation gravitational wave 
detectors such as the Einstein
Telescope\footnote{http://www.ego-gw.it/ILIAS-GW/FP7-DS/fp7-DS.htm} 
 (currently at the study stage) enter service, in
conjunction with future X-ray timing missions\footnote{Such as the proposed
High Time Resolution Spectrometer (HTRS) on the International X-ray
Observatory (IXO), or the
Advanced X-ray Timing Array (AXTAR, \citealt{Chakrabarty08b})}, there is a
very strong possibility of constraining pulsar torque models in this
way.  

Having some way of timing spin
in quiescence would also be useful: there has been a suggestion that
at least one of the accreting 
millisecond pulsars, SAX J1808.4-3658, turns on as a radio pulsar in quiescence
\citep{Burderi03, Campana04}.  Searches for radio pulsations from this
source at 1.4 GHz have so far been unsuccessful \citep{Gaensler99,
  Burgay03}, as have searches
at 2.0 and 4.6 GHz
with the Green Bank Telescope (Adrienne Juett, private
communication).  Nonetheless future radio telescopes such as LOFAR, the Murchison
Wide Field Array, and eventually the Square Kilometer Array may
improve detection prospects.   

\section{Conclusions}

With two exceptions \citep{Abbott07, Abbott07a}, most of the effort in
periodic gravitational wave searches to date has gone into searches
for signals from isolated non-axisymmetric neutron stars.  These
include both targeted searches for known pulsars, and blind searches
for unknown systems.  While these blind surveys are computationally
limited, searches for accreting neutron stars can be even more
challenging in the absence of well-defined orbital parameters. Given
the computational resources required, it is important that the
searches are well motivated from an astrophysical standpoint.  For
isolated neutron stars, the spindown rate provides an indirect limit
by assuming that all of the spindown is due to gravitational wave
emission.  However, the true amplitude is almost certainly much
smaller than this; most of the spindown is well explained by
electromagnetic braking (see e.g. \cite{Palomba00} for estimates).
For accreting neutron stars the indirect limit is provided by the spin
balance argument and, if nature does employ this mechanism, the true
amplitude is likely to be very close to this limit.  Thus, the first
direct detection of periodic gravitational radiation might very well
be from accreting neutron stars.  This is, however, going to be
challenging, given the current state of our
knowledge regarding spin and orbital parameters.  It may be possible
to reduce this uncertainty, but otherwise some out
of the box thinking will be required.

One area that may lead to progress is a more in-depth consideration of
prospects for indirect
detection.  Ideally what one would like is a signature of one
of the gravitational wave emission mechanisms that could be detected
more easily in
electromagnetic wavebands.  One possibility is to look for the
signature of the mechanism that generates the
quadrupole. \citet{Ushomirsky00}, for example, point out that a
quadrupole induced by temperature asymmetries in the deep crust may
lead to detectable asymmetries in surface X-ray emission, provided that the
gradients do not wash out as radiation escapes through the ocean and
atmosphere. \citet{Payne06b} have also considered the implications of a
magnetic mountain on the observable properties of Type I X-ray bursts
and burst oscillations.   Ideas along these lines merit more detailed study.

A second possibility for indirect detection is to look for the effect of
angular momentum loss associated with gravitational wave emission by
monitoring spin evolution.   As pointed out by \citet{Chakrabarty08},
X-ray timing (for accreting millisecond pulsars) is able to detect
very small changes in spin rate, and is potentially
much more sensitive to gravitational wave spin-down than
the gravitational wave detectors themselves. Measuring spin variations
in outburst is complicated by uncertainties associated with pulse
profile variability, but spin changes between outbursts may provide
stronger constraints.  The main issue is then whether it is possible
to make an unambiguous identification of the cause of the spin-down,
given the other negative torque mechanisms that may operate.  

We may also want to consider a change of emphasis in our theoretical
modeling.   Rather than focusing on what might be detectable, theorists
should perhaps consider what upper limits 
would be most physically interesting.  In the spin balance scenario
there are several sources that 
could be detectable by Advanced LIGO if we searched smaller
frequency bands:  so which bands should we search?  The strongest limits on
gravitational wave amplitude will be obtained if we
restrict searches to a low frequency band around the sweet spots of
the detectors.  The limits on quadrupole moment $Q$, however, will be tighter if
we focus on a higher frequency band.  Which of these options would
make the biggest difference to our understanding of neutron star
physics or binary evolution?  Is it better to search
at the dominant r-mode frequency (four thirds the spin
frequency) or the fixed quadrupole frequency (twice the spin
frequency)?   Should we be focusing our attention on spin-down
scenarios rather than spin-balance scenarios?  Certainly for the
accreting millisecond pulsars, where simultaneous X-ray timing may
enable spin tracking, this latter scenario is much more attractive and
(given current estimates of spin variation) could lead to physically
meaningful constraints on torque models.   We cannot hope to answer
all of these questions in such a short paper.  However these are issues
that the community should be debating if we are to make the most of
the data that 
the gravitational wave observatories obtain in the near future.

\section{Acknowledgments}

ALW thanks COSPAR and the Leids Kerkhoven-Bosscha Fonds
(LKBF) for financial support to attend the 37th
COSPAR meeting in Montreal. We would also like to thank Lars Bildsten,
Deepto Chakrabarty, Jason Hessels, Adrienne Juettt and Bernard Schutz
for useful discussions that informed this work.

\bibliographystyle{elsart-harv}
\bibliography{mont_gw}   

\end{document}